\documentclass[twocolumn]{aastex61} 

\received{XX 1, 2017}
\revised{XX 1, 2017}
\accepted{\today}
\submitjournal{ApJ}

\shorttitle{The distribution and ages of star clusters in the Small Magellanic Cloud}
\shortauthors{Bitsakis et al.}

\begin{document}

\title{The distribution and ages of star clusters in the Small Magellanic Cloud: Constraints on the interaction history of the Magellanic Clouds}

\correspondingauthor{Theodoros Bitsakis}
\email{t.bitsakis@irya.unam.mx}

\author{Theodoros Bitsakis}
\affiliation{Instituto de Radioastronom\'ia y Astrof\'isica, Universidad Nacional Aut\'onoma de M\'exico, Morelia, 58089, Mexico}

\author{R. A. Gonz\'alez-L\'opezlira}
\affiliation{Instituto de Radioastronom\'ia y Astrof\'isica, Universidad Nacional Aut\'onoma de M\'exico, Morelia, 58089, Mexico}

\author{P. Bonfini}
\affiliation{Instituto de Radioastronom\'ia y Astrof\'isica, Universidad Nacional Aut\'onoma de M\'exico, Morelia, 58089, Mexico}

\author{G. Bruzual}
\affiliation{Instituto de Radioastronom\'ia y Astrof\'isica, Universidad Nacional Aut\'onoma de M\'exico, Morelia, 58089, Mexico}
 
\author{G. Maravelias}
\affiliation{Instituto de F\'isica y Astronom\'ia, Universidad de Valpara\'iso, Valpara\'iso, Chile}
\affiliation{Department of Physics, University of Crete, GR-71003, Heraklion, Greece}

\author{D. Zaritsky}
\affiliation{Steward Observatory, University of Arizona, Tucson, AZ 85719, USA}

\author{S. Charlot} 
\affiliation{Sorbonne Universit\'es, UPMC-CNRS, UMR7095, Institut d'Astrophysique de Paris, F-75014 Paris, France}

\author{V. H. Ram\'irez-Siordia}
\affiliation{Instituto de Radioastronom\'ia y Astrof\'isica, Universidad Nacional Aut\'onoma de M\'exico, Morelia, 58089, Mexico}



\begin{abstract}

We present a new study of the spatial distribution and ages of the star clusters in the Small Magellanic Cloud (SMC). To detect and estimate the ages of the star clusters we rely on the new fully-automated method developed by \citet{Bitsakis17}. Our code detects 1319 star clusters in the central 18 deg$^{2}$ of the SMC we surveyed (1108 of which have never been reported before). The age distribution of those clusters suggests enhanced cluster formation around 240 Myr ago. It also implies significant differences in the cluster distribution of  the bar with respect to the rest of the galaxy, with the younger clusters being predominantly located in the bar. Having used the same set-up, and data from the same surveys as for our previous study of the LMC, we are able to robustly compare the cluster properties between the two galaxies. Our results suggest that the bulk of the clusters in both galaxies were formed approximately 300 Myr ago, probably during a direct collision between the two galaxies. On the other hand, the locations of the young ($\le$50 Myr) clusters in both Magellanic Clouds, found where their bars join the HI arms, suggest that cluster formation in those regions is a result of internal dynamical processes.  Finally, we discuss the potential causes of the apparent outside-in quenching of cluster formation that we observe in the SMC. Our findings are consistent with an evolutionary scheme where the interactions between the Magellanic Clouds constitute the major mechanism driving their overall evolution.  

\end{abstract}






\section{Introduction}
The Magellanic Clouds have significantly advanced our understanding on galaxy evolution. Owing to their proximity, individual stars can be observed, providing important information about the spatially resolved star formation, and the origin and properties of their stellar populations. 

The Small Magellanic Cloud (SMC) is a dwarf irregular galaxy located at a distance of $\sim$60.6 kpc \citep{Hilditch05}. Simulations supported by observational evidence suggest that it evolved in tandem with its counterpart -- the Large Magellanic Cloud (LMC), thus sharing a common interaction and star formation history \citep[e.g., see][and references therein]{Besla12}. \citet{Yoshizawa03} performed N-body simulations of the tidal distortions and concluded that the two galaxies should have interacted over the past $\sim$0.2 Gyr. Their results are partially supported by \citet{Harris04}, who studied the spatially resolved star formation history of the SMC and showed that it underwent various periods of enhanced star formation $\sim$2.5, 0.4, and 0.06 Gyr ago. They are also in agreement with \citet{Chiosi06} and \citet{Glatt10}, who suggested that the close interaction between the two Clouds have resulted in the triggering of cluster formation activity.

More recently, \citet{Besla07} and \citet{Kallivayalil13} challenged the scenarios where the Magellanic Clouds have already completed several orbits around the Galaxy, using current {\it Hubble Space Telescope} (HST) proper motion measurements; they suggested that the Clouds are in their first orbit passage about the Galaxy. Moreover, \citet{Besla12} studied the interaction history of those galaxies using numerical models constrained by the HST observations and showed that, while they have not interacted before with the Galaxy, the Magellanic Clouds must have experienced a direct collision some time 100-to-300 Myr ago. This seems to agree with the findings of \citet{Harris07}, who studied the stellar populations of the Magellanic Bridge -- the tidal stream of neutral gas and stars possibly associated with the interaction of the two galaxies -- and showed that the star formation in the Bridge commenced some time 200-300 Myr ago. A direct cloud-cloud collision would also explain the existence of a small population of SMC stars -- based on their peculiar kinematics and metallicities -- which were found in the LMC \citep{Olsen11}. In spite of all this progress, the question of whether the evolution of the Magellanic Clouds is driven by internal processes (i.e., the action of bars, morphological/dynamical quenching) or environmental mechanisms (i.e., galaxy interactions) is still unclear. One would expect that in the case of environmental evolution many of the properties of the two galaxies (e.g., the star formation history) would be correlated.

A robust method to explore the formation and interaction histories of nearby galaxies, where individual stars can be resolved (such as the Magellanic Clouds), entails the study of the age distribution of their star clusters. Owing to modern instrumentation which allows us to estimate their ages and metallicities with high precision -- in contrast with field stars -- star clusters represent unique tools to constrain the star formation history of their host galaxies and to disentangle the special conditions they might have undergone. Despite the plethora of studies of the star clusters in the Magellanic Clouds, the lack of a statistically robust detection method that creates uniform and complete samples (as opposed to the visual identification methods that are usually applied) has posed significant limitations for the systematic study of the star cluster formation history of both galaxies. In \citet{Bitsakis17}, we presented a new fully-automated method to robustly detect and estimate the ages of star clusters in nearby galaxies. Using statistical analysis on high resolution maps of the LMC, we obtained a large, uniform sample of star clusters (in the central 49 deg$^{2}$) which we exploited to put constraints on the formation history of that galaxy. A similar analysis is followed in the current study using the same method and data surveys for the SMC. In Section 2, we describe the dataset we use in the current study. Section 3 contains a brief description of the cluster detection and age estimation codes (a more analytic description along with statistical tests can be found in \citealt{Bitsakis17}). The results are presented in Section 4, while in Section 5 we make a comparison of the SMC-LMC star cluster age distributions and derive useful conclusions about their interaction history. Finally, in Section 6 we summarize our findings. 

Throughout this work we assume a distance modulus to the SMC of 18.91 mag \citep{Hilditch05}.

\begin{figure*}
\begin{center}
\includegraphics[scale=0.3]{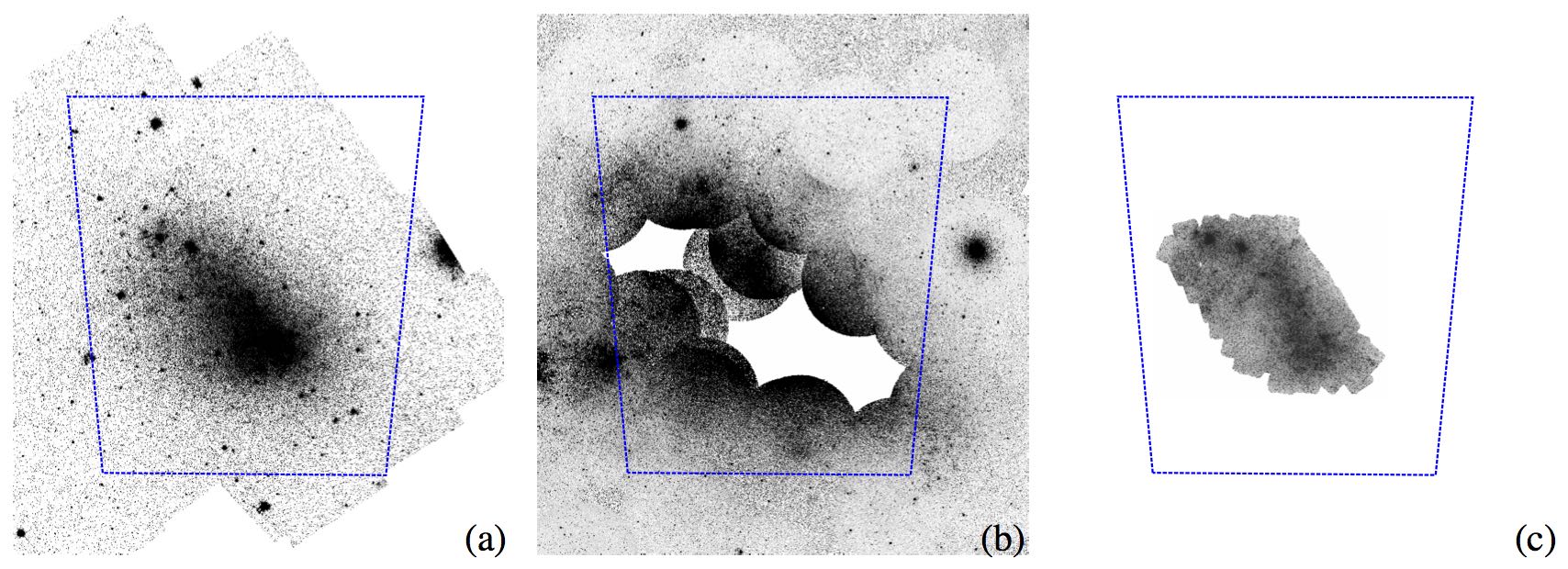}
\caption{(a) The Spitzer/IRAC 3.6$\micron$ \citep{Gordon11}, (b) the GALEX/NUV \citep{Simons14}, and (c) the SWIFT/UVOT \citet{Siegel14} mosaics of the SMC, respectively. The dashed blue box indicates the area covered by MCPS \citep{Zaritsky02}, which was also surveyed by our code. }
\label{mcps_coverage}
\end{center}
\end{figure*}

\section{The data}
We have made use of archival data of the SMC at various bands. \citet{Simons14} presented the near-ultraviolet mosaic ($\lambda_{\rm eff}$=2275\AA) of that galaxy obtained by the Galaxy Evolution Explorer \citep[GALEX;][]{Martin05}. The median exposure time was 733 seconds, and the 5$\sigma$ depth of point sources varied between 20.8 and 22.7 mags. Although the mosaic covers a region of 63 deg$^{2}$, which contains the SMC bar, wing and tail, there are two sub-regions that were not observed, of $\sim$0.25 and 1 deg diameter, north east and south west from the center, respectively (see Figure~\ref{mcps_coverage}b). These holes in the coverage were compensated for with the Swift Ultraviolet-Optical Telescope (UVOT) Magellanic Clouds Survey \citep[SUMAC;][]{Siegel14}, which imaged the central 3.8 deg$^{2}$ of the galaxy (Figure~\ref{mcps_coverage}c) with deeper exposures of 3000 s in all three $NUV$ filters of the instrument ($UVW1$, $UVW2$, and $UVM2$).

Our infrared data come from the ``Surveying the Agents of a Galaxy's Evolution SMC survey'' \citep[SAGE-SMC;][]{Gordon11}, that mapped the full SMC (30 deg$^{2}$) with both the Infrared Array Camera \citep[IRAC, Figure~\ref{mcps_coverage}a; ][]{Fazio04} and the Multiband Imaging Photometer \citep[MIPS;][]{Rieke04} on-board the Spitzer Space Telescope. It produced mosaics at 3.6, 4.5, 5.8, and 8.0$\micron$ with IRAC, and at 24, 70, and 160$\micron$ with MIPS, with integrated exposure times of 63 hours in the IRAC and $\sim$400 hours in the MIPS bands, respectively.  

Finally, we exploited the photometric information by \citet{Zaritsky02}, who presented the stellar catalog and extinction map of the SMC, as part of the Magellanic Cloud Photometric Survey (MCPS; marked with dashed blue lines in Figure~\ref{mcps_coverage}). They obtained 3.8-5.2 minute exposures of the central 18 deg$^{2}$ of the SMC in the Johnson $U$, $B$, $V$, and Gunn $i$ bands with the Las Campanas Swope Telescope under 1.5 arcsecond seeing conditions. The  limiting magnitudes varied, depending on the filter, between 21.5 mag for $U$ and 23.0 mag for $i$. Using DAOPHOT II \citep{Stetson87}, they created a photometric catalog that contains 24.5 million sources in all the area covered by the MCPS (including the SMC, LMC and the Magellanic Bridge). They also estimated the line-of-sight extinctions to the stars in their catalog and  produced an extinction map of the SMC. This was achieved by comparing the observed stellar colors with those derived from the stellar photospheric models of \citet{Lejeune97}. Thus they estimated the effective temperature ($T_{\rm eff}$) and measured the extinction ($A_{V}$) along the line of sight to each star, adopting a standard Galactic extinction curve. They produced two $A_{V}$ maps, one for hot (12000 K $<$ $T_{\rm eff}$ $\le$ 45000 K) and one for cool (5500 K $<$ $T_{\rm eff}$ $\le$ 6500 K) stars. In Figure~\ref{mcps_coverage}, we present the coverage of MCPS in comparison with that of other surveys we used for the detection of the star clusters; one can see that the central 18 deg$^{2}$ of the SMC are imaged.

\section{The cluster detection and age estimation method}

\begin{table*}
\begin{minipage}{120mm}
\begin{center}
\caption{SMC star cluster catalog.}
\label{tab_clusters}
\begin{tabular}{lccccccc}
\hline 
\hline
 & R.A.(J2000) & Dec(J2000) & Radius & log(Age) & Lower unc. & Upper unc. & Bica et al. (2008)  \\
 catalog ID   & (deg)         & (deg)            & (deg)    & (yr) & (yr) & (yr) &   catalog ID \\
\hline
  SMC-NUV-484     &  14.0765 &  -72.4634 &  0.0280  &   7.22 &  6.92 &  7.33  &  343 \\
  SMC-M2-287 &   13.0482 &  -72.5310  &  0.0065   & 7.99  & 7.77  & 8.01 &  258  \\
  SMC-IR1-449  &    13.3365 &  -73.1764 &  0.0130  &   7.96 &  7.88 &  8.06  &  -- \\
  ... & ... & ... & ... & ... & ... & ... \\
\hline
\end{tabular}
\end{center}
{\bf Notes. } \\
The lower and upper uncertainty bounds are estimated at the 16$^{\rm th}$ and 84$^{\rm th}$ percentiles, respectively. (The full version is available online.)\\ 
\end{minipage}
\end{table*}%

The code we used here to automatically detect and estimate the ages of the SMC star clusters was analytically described in \citet{Bitsakis17}. Summarizing, the code makes use of the star counts method \citep[see][and references therein]{Schemja10}, which estimates the density of stars in a given region-of-interest and finds overdensities above some local background threshold ($\Sigma_{\rm det}$). To define the relation between $\Sigma_{\rm det}$ and the background density we performed Monte-Carlo simulations with artificial star clusters, having both Gaussian as well as uniform overdensity profiles (accounting for both compact and diffuse clusters), projected over various background values. The code is applied on a pixel-map conversion of the original image, where each star is represented by a single pixel. Only stars located in the overdensities are considered and a source detection is applied on the smoothed final image to define the center and radius of each candidate cluster. The method has been proven to be fast and accurate and was initially tested on the LMC with impressive results \citep[see][]{Bitsakis17}, yielding the discovery of 3500 new star clusters that have never been reported before. For the sake of consistency we use the same setup as for the LMC; we run the detection sequence on the ultraviolet (GALEX/NUV, SWIFT/UVM2) and near-infrared mosaics (Spitzer/IRAC 3.6) of the SMC in order to probe different cluster ages (e.g. young clusters are expected to host massive UV-emitting stars, while old clusters are dominated by low-mass stars emitting mostly in the near-IR part of the spectrum). We then use the MCPS catalog to obtain the photometric information of the stellar populations. The detection sequence yields a total of 2219 \emph{candidate} clusters and associations in the corresponding region. 

\begin{figure*}
\begin{center}
\includegraphics[scale=0.4]{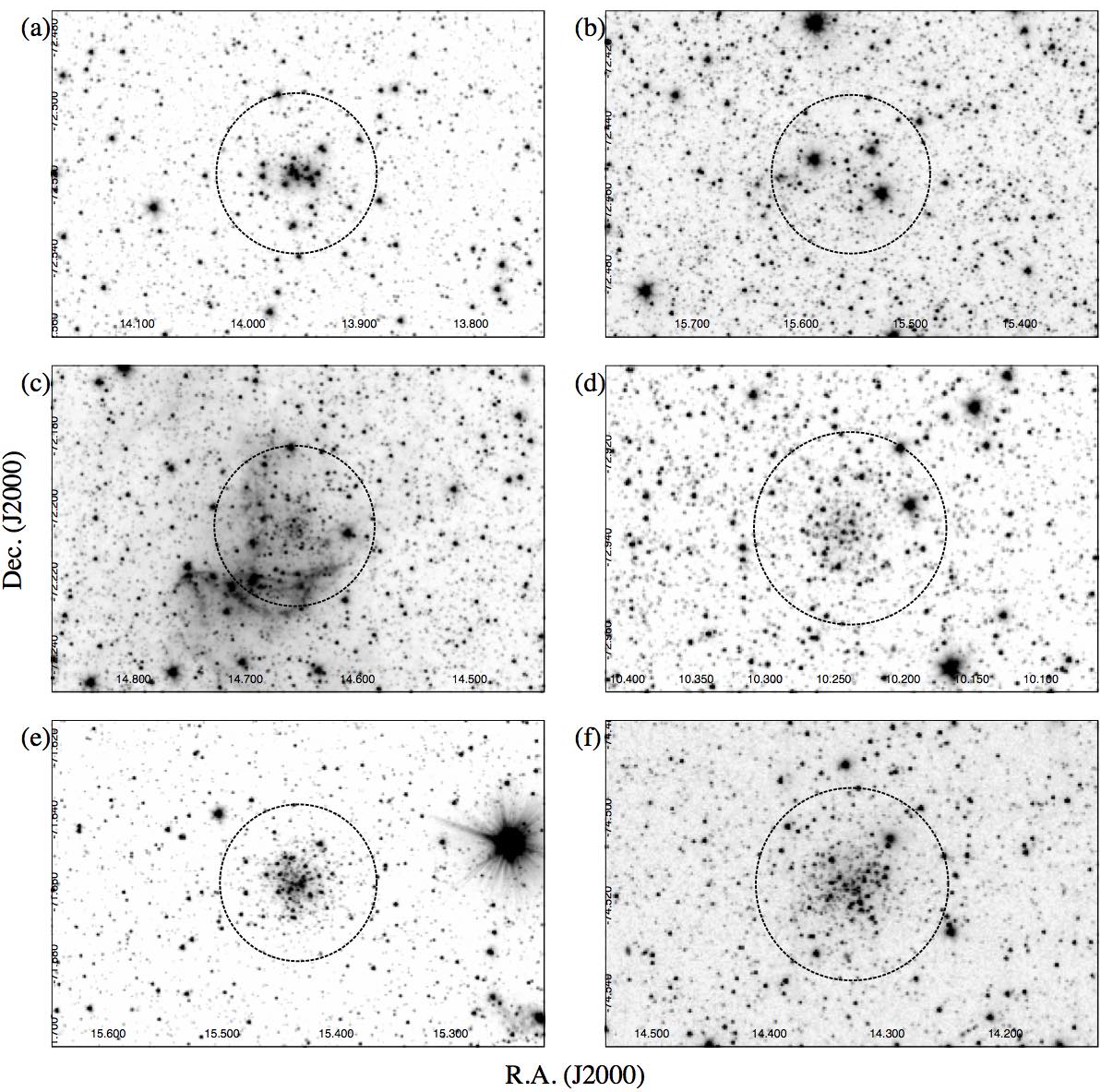}
\caption{Examples of clusters from our catalog presented on the Spitzer $IRAC~3.6\micron$ image. The dashed black lines mark the radii, as defined by the star-counts code. (a): cluster SMC-NUV-484, age 16.6$^{+8.2}_{-4.7}$ Myr; (b) SMC-IR1-665, age 48.5$^{+2.9}_{-2.0}$ Myr; (c) SMC-IR1-635, age 186$^{+55}_{-35}$ Myr; (d) SMC-IR1-358, age 512$^{+135}_{-124}$ Myr; (e) SMC-IR1-727, age 845$^{+284}_{-650}$ Myr; and (f) SMC-IR1-270, age 1.07$^{+0.23}_{-0.85}$ Gyr. The horizontal and vertical axes (plural) show, respectively, correspond to R.A. and Declination measured in degrees (J2000). }
\label{fig_clusters}
\end{center}
\end{figure*}

The age estimation algorithm (also presented in \citealt{Bitsakis17}) consists of a modified version of the code of Ram\'irez-Siordia et al. (in prep.). Briefly, this code uses a Bayesian approach to obtain the most likely theoretical isochrone that reproduces the observed CMD of each candidate cluster, while taking into account the cluster star memberships. The set of 80 model isochrones we used here is a byproduct of an independent project by Charlot \& Bruzual (in preparation)\footnote{The Charlot \& Bruzual isochrones are available to the interested user upon request.}, and was produced following the evolutionary tracks of \citet{Chen15} and accounting for the evolution of thermally pulsing asymptotic giant branch (TP-AGB) stars \citep{Marigo13}. The isochrones were calculated for a representative SMC metallicity of [Fe/H]=-0.70 \citep[i.e. $Z$=0.004;][]{Venn99}, and cover the range 6.9 $\le$ log(age) $<$ 9.7 yr.

As anticipated above, we also perform field star decontamination. Our code uses a modified version of the method described in \citet{Mighell96}. According to this, the code produces the CMD of the candidate cluster as well of its surrounding field stars and estimates the probability of each candidate star to belong to the cluster. This membership probability is stored in a table containing all the cluster star information and is eventually used during the age estimation process mentioned above. In \citet{Bitsakis17} we showed that the method performs well even in high field star density environments (such as the LMC/SMC bar). Eventually, the code discards any candidate cluster with an insignificant number of stars ($n<$20) having high membership probability ($>$60\%), as well as those clusters that could not be fitted by our age estimation code. 

\begin{figure*}
\begin{center}
\includegraphics[scale=0.4]{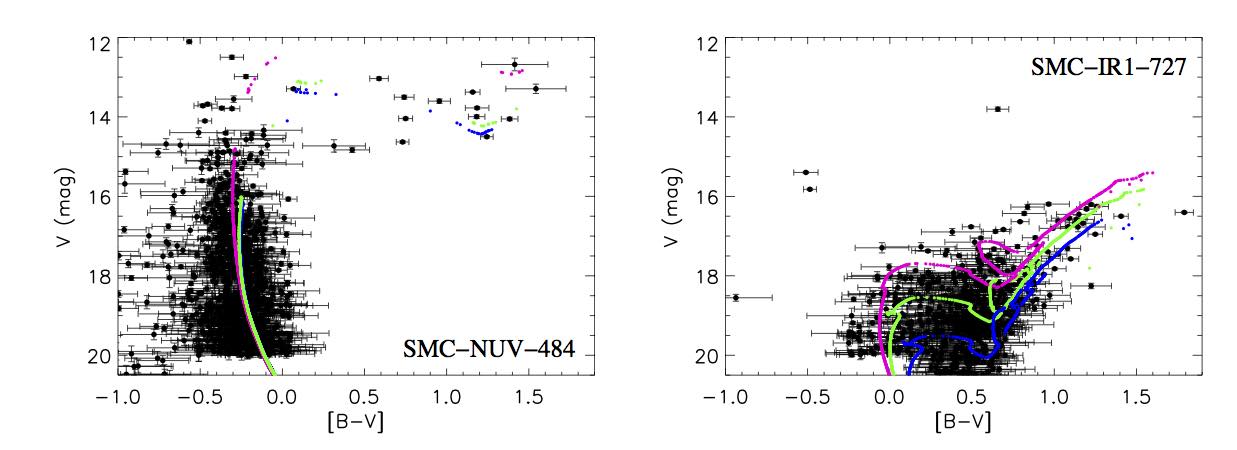}
\caption{Examples of the isochrone fitting process in the ($B-V$) versus $V$ field star decontaminated CMDs of the star clusters SMC-NUV-484 and SMC-IR1-727, presented in Fig.~\ref{fig_clusters}. Best fit isochrones are presented in green, upper and lower uncertainties in magenta and blue, respectively. }
\label{fig_isoc}
\end{center}
\end{figure*}

To ensure a more accurate age estimation we perform the CMD fitting in the ($U-V$) versus $V$, ($B-V$) versus $V$, and ($V-i$) versus $i$ bands for each cluster and then we combine the final results using equation 5 from \citet{Bitsakis17}, which takes into account the number of stars included, and how well the age is constrained in each fitting. In Figure~\ref{fig_isoc}, we present two examples of the best age estimation in the CMDs of clusters SMC-NUV-484 and SMC-IR1-727. The final catalog contains 1319 \emph{secure} clusters (40\% smaller than the initial \emph{candidate} cluster sample).  These clusters are presented in Table~\ref{tab_clusters}; column (1) gives the cluster identifier (it consists of a reference to the band where each cluster was initially detected, i.e., $IR1$ refers to Spitzer/IRAC1, $NUV$ to GALEX/NUV, and $M2$ to SWIFT/UVM2, plus the serial number of the corresponding cluster); columns (2) and (3), respectively, contain the right ascension (R.A.) and declination (Dec.) of the cluster centers, in J2000 decimal equatorial coordinates; column (4) reports the cluster radii; columns (5), (6), and (7) contain, respectively, the best age estimation for each cluster, and its lower and upper uncertainty bounds (derived from the 16$^{\rm th}$ and 84$^{\rm th}$ percentiles of the probability distribution histogram produced by the code). Finally, column (9) contains -- if available -- the corresponding cluster identifier from the catalog of \citet{Bica08}. Some characteristic examples of clusters ordered by increasing age are presented in Figure~\ref{fig_clusters}.

\section{Results}

\begin{figure*}
\begin{center}
\includegraphics[scale=0.64]{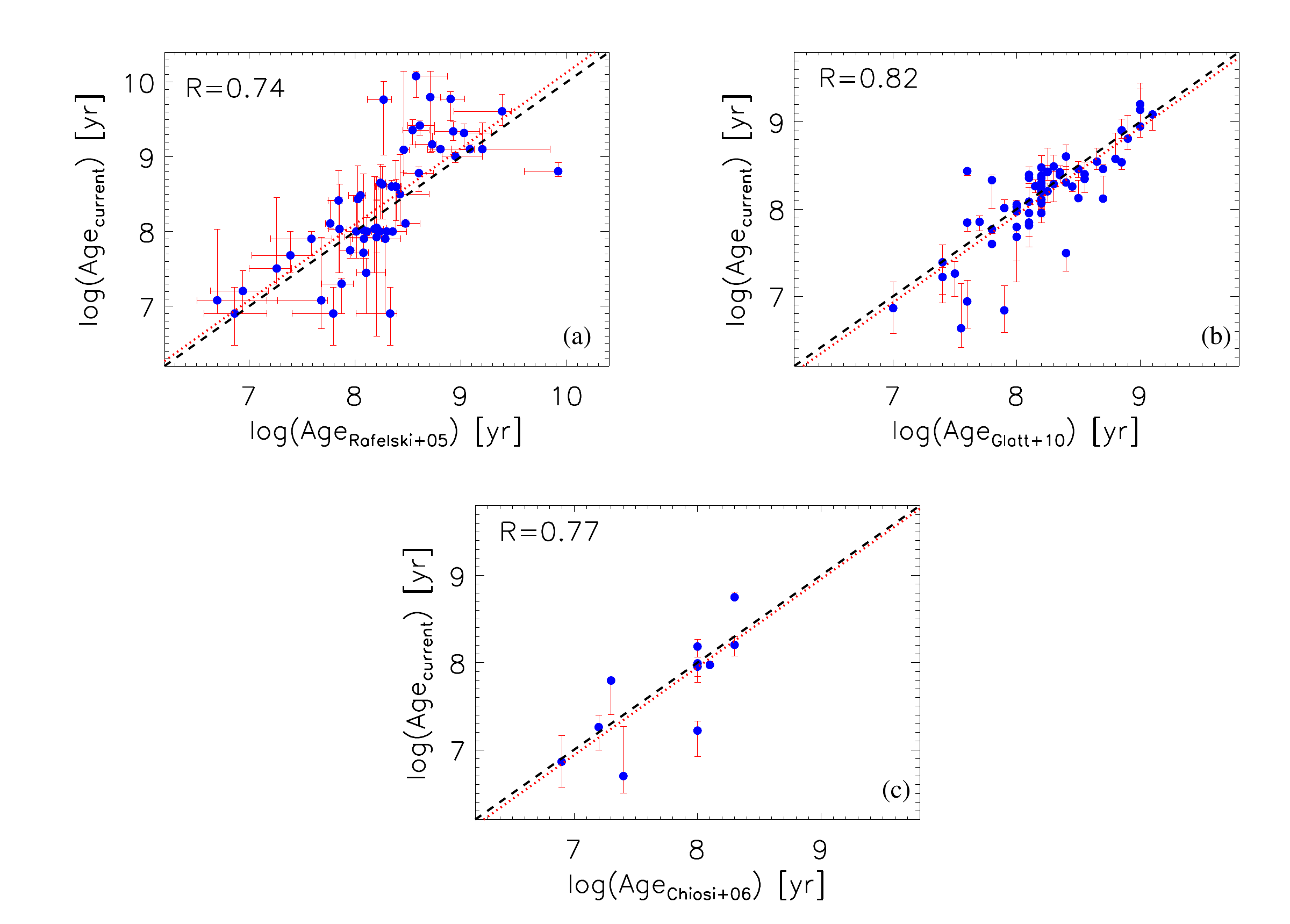}
\caption{Comparison of the ages determined from our method ($\rm Age_{\rm current}$) for clusters we have in common with (a) \citet{Rafelski05}, (b) \citet{Glatt10}, and (c) \citet{Chiosi06}. The dashed black lines correspond to the one-to-one correlation, while the dotted red ones are the least square fits to the data. The Pearson correlation coefficients (R) are indicated in the upper left corner of each panel.}
\label{fig_comp_ages}
\end{center}
\end{figure*}

\subsection{Comparisons with other surveys}
We compare our final catalog of star clusters with that of \citet{Bica08}. These authors have reported 515 clusters in the central 18 deg$^{2}$ of the SMC we surveyed, 211 of which (58\%) overlap our sample. In Figure~\ref{fig_comp_ages}, we compare our age estimates with those from other surveys. \citet{Rafelski05} compared the integrated colors of their star clusters, acquired from the MCPS survey, with models of simple stellar populations. Unfortunately, their technique is not able to decontaminate from field stars; hence, although these authors performed various tests to ensure the reliability of their estimates, their method can introduce significant biases, especially at high field star density regions (like the SMC bar). Thus, the comparison with their results yields a Pearson R-coefficient 0.74 (see also Figure~\ref{fig_comp_ages}a). On the other hand, \citet{Glatt10} visually fitted a set of isochrone models to the observed cluster CMDs. Although they used a field star decontamination technique, the large uncertainties introduced by visual identification of the main sequence turn-off are likely the origin of the large scatter between theirs and our age estimates, having R=0.82 (see Figure~\ref{fig_comp_ages}b). Similarly, \citet{Chiosi06} corrected for field star contamination, and used both visual and $\chi^{2}$ minimization methods; they divided the observed and model CMDs in bins of color and magnitude, and minimized their differences. Although we only have 11 clusters in common, the comparison yields R=0.77 (see Figure~\ref{fig_comp_ages}c). Finally, \citet{Parisi14} carefully calculated the ages of a small sample of 15 old SMC clusters using high spatial resolution data from the Very Large Telescope in Chile. For the only cluster we have in common (identified as L17 in their catalog, our SMC-IR1-226), we measure an age 1.22$^{+0.11}_{-0.40}$ Gyr, which is remarkably similar to their 1.25 Gyr estimate.


\subsection{The age distribution of star clusters}

\begin{figure}
\begin{center}
\includegraphics[scale=0.5]{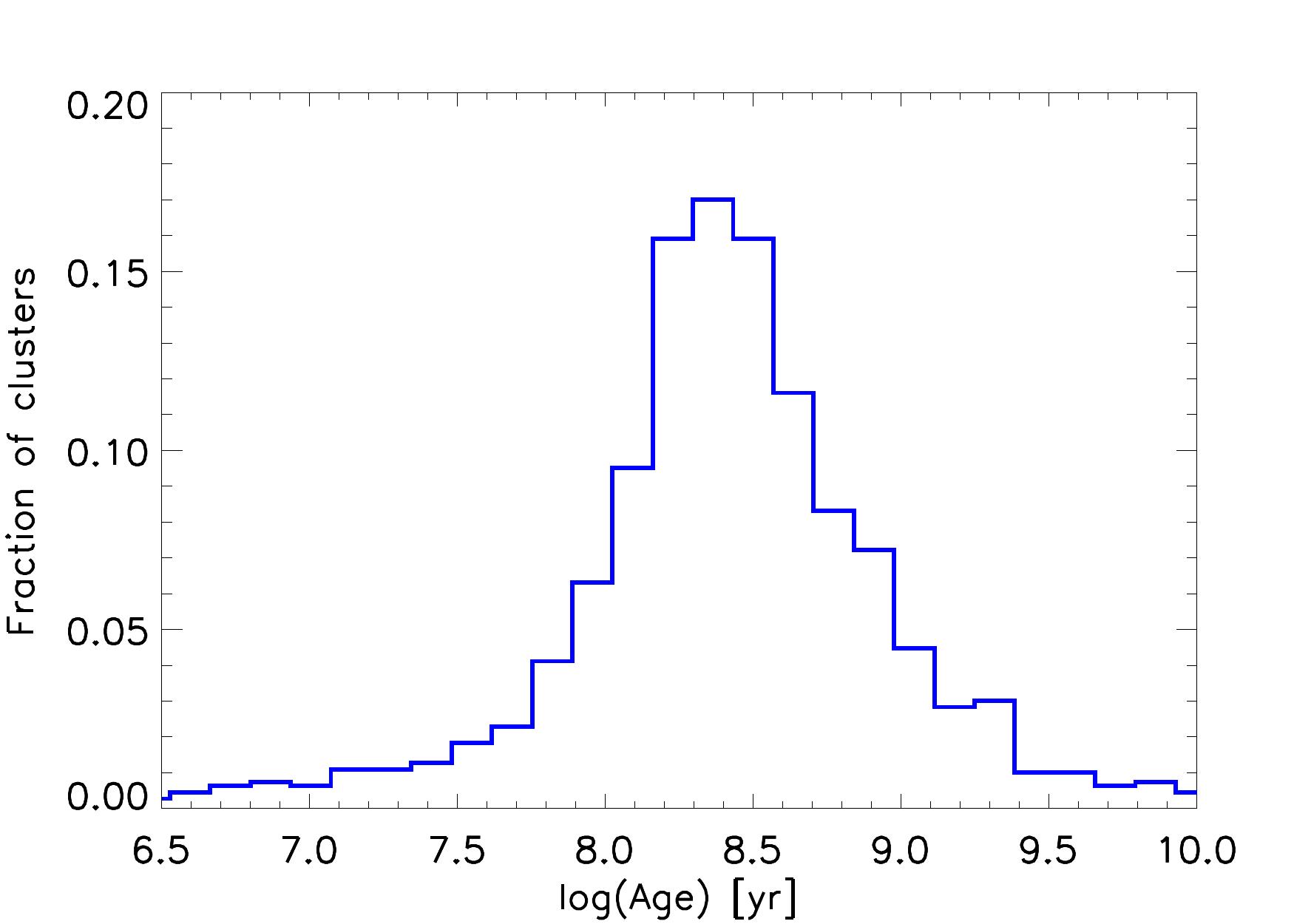}
\caption{Age distribution of the SMC clusters. The fractions presented here are normalized to the total number of clusters found in that galaxy.} 
\label{fig_hist_ages}
\end{center}
\end{figure}

In Figure~\ref{fig_hist_ages}, we present the age distribution of star clusters in the SMC. The bin size was optimized using the Freedman-Diaconis rule (bin size 0.136 dex). The main cluster formation event seems to have happened $\sim$240 Myr ago. The decline in the number of star clusters beyond the main peak could be associated both with cluster fading \citep[e.g.,][]{Boutloukos03}, and/or cluster dissolution due to a variety of mechanisms, such as $(i)$ residual gas expulsion, $(ii)$ two-body relaxation, $(iii)$ tidal heating from disc shocks, and $(iv)$ tidal harassment from giant molecular clouds \citep[see][and references therein]{Baumgardt13}. On the other hand, phenomena like the cluster disruption due to gas expulsion after the burst of star formation took place in the initial stages of cluster formation, and therefore in short time-scales \citep[$\sim$40 Myr for the Magellanic Clouds; see][]{deGrijs09}.

\begin{figure}
\begin{center}
\includegraphics[scale=0.5]{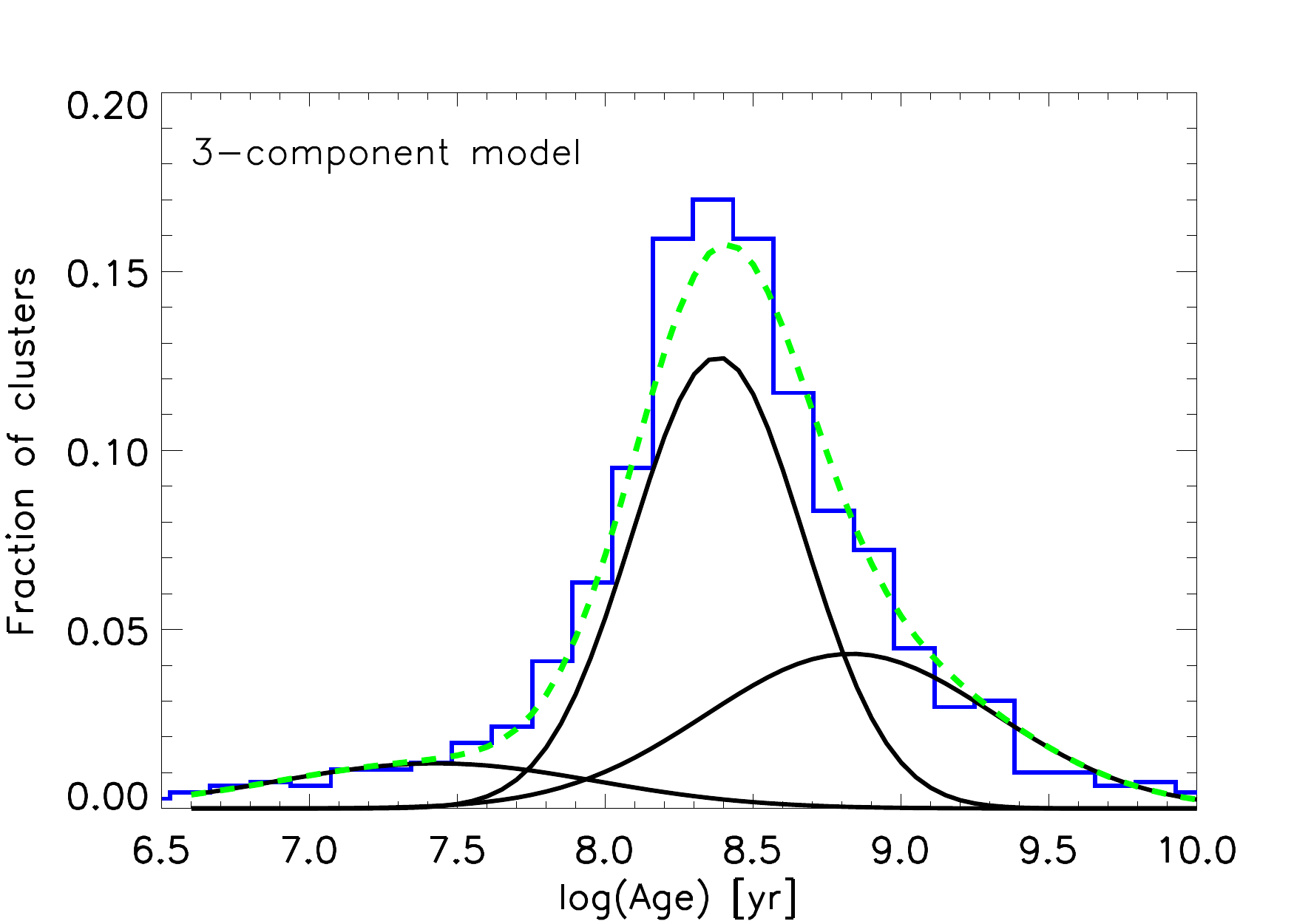}
\caption{The three component mixture model (dashed green line), and its individual constituents (solid black lines). The fractions presented here are normalized to the total number of clusters found in that galaxy.}
\label{fig_mix}
\end{center}
\end{figure}

\begin{figure*}
\begin{center}
\includegraphics[scale=0.7]{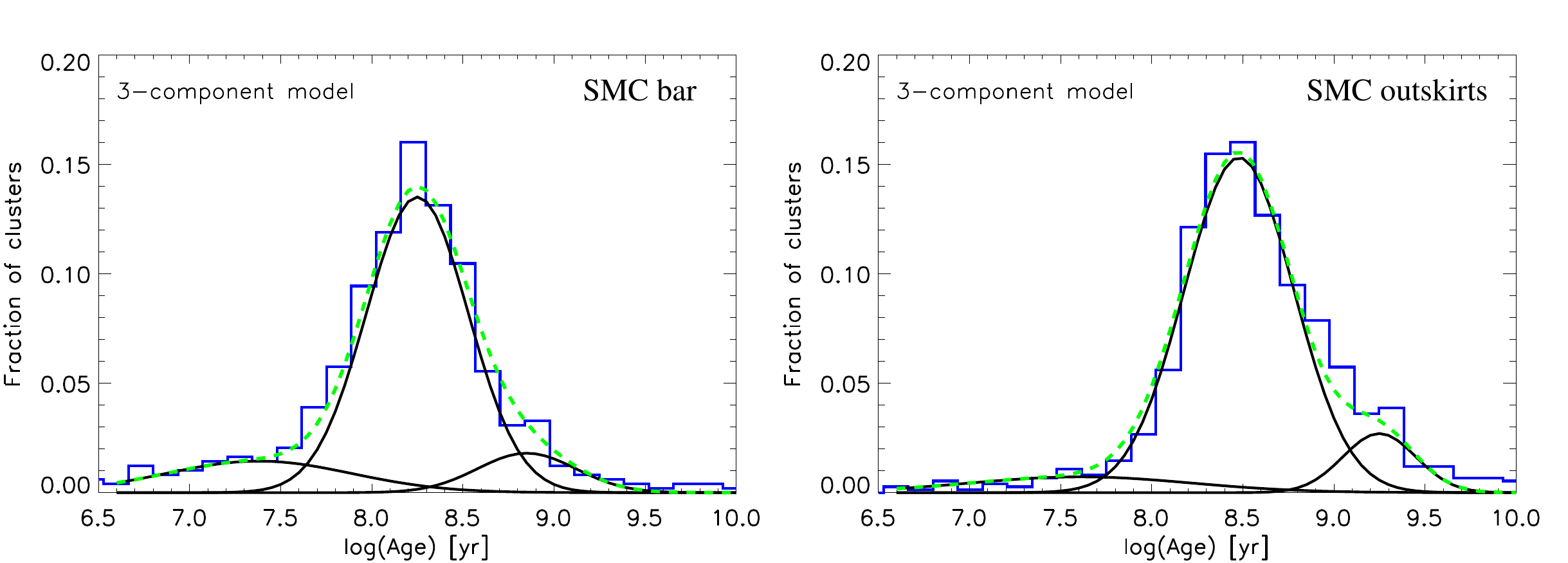}
\caption{Age distributions of the star clusters found in the SMC bar (left panel) and in the rest of the galaxy (right panel). We also display the three component mixture models in both figures (dashed green lines), and their individual constituents (solid black lines). The fractions presented here are normalized to the total number of clusters found in that galaxy.}
\label{fig_hist_ages_bar}
\end{center}
\end{figure*}

Since a star cluster formation event in our data could be represented by a single Gaussian distribution (due to the range of uncertainties in the estimation of the cluster ages), we use a Gaussian mixture model code, NMIX\footnote{Publicly available at \url{https://people.maths.bris.ac.uk/~mapjg/Nmix.}}, to derive the underlying number of such distributions in our data. This method reports the statistically motivated number of Gaussian distributions that can fit a given dataset by implementing the approach of \citet{Richardson97}. In Figure~\ref{fig_mix}, we present the results of the fitting; it is shown that our cluster age distribution can be successfully reproduced by a three component mixture model (having Bayes K-factors between that model and each one of the rejected univariate distributions $>$4.5), with peaks 30, 240, and 680 Myr ago. Based on their results, \citet{Glatt10} have visually identified and proposed two main periods of cluster formation 160 and 630 Myr ago, as well as a minor event $\sim$50 Myr ago (see Figure 5 of that work); this last event of star formation was also detected by \citet{Harris04}. Whereas the 50 and 630 Myr peaks from \citet{Glatt10} are consistent with our secondary cluster formation events, the 160 Myr one is significantly different from our main 240 Myr event. We note here that histogram peaks can be also the result of binning artifacts. This is not the case for our findings since NMIX fits models on the un-binned data. To test whether binning could be at the origin of the discrepancy with \citet{Glatt10}, we applied the Freedman-Diaconis rule to calculate the bin size for their sample; its value is 0.109 dex. Using this bin size, we produced an updated version of the \citet{Glatt10} histogram, which shows a major formation event 280 Myr ago, with minor ones appearing 20, 100, and 450 Myr ago. This exercise suggests that, in addition to the scatter mentioned in \S4.1, differences in the binning scheme also contribute to the different results obtained by \citet{Glatt10} and in the present work.

\subsection{The spatial age distribution of star clusters}
To further study the cluster formation history in the SMC, we present in the two panels of  Figure~\ref{fig_hist_ages_bar} the age distributions of those clusters located in the bar (left panel) and everywhere else in the galaxy (hereafter referred to as ``outskirts''; right panel). The two distributions display important differences, having a Kolmogorov-Smirnov probability of being drawn from the same sample $<$10$^{-5}$. In contrast to the bar that had a major formation event around 200 Myr ago, with secondary peaks appearing at 20 and $\sim$800 Myr,  the outskirts' major peak appeared $\sim$270 Myr ago, with secondary ones 40 Myr and 2 Gyr ago. These results are drawn from the 3-component NMIX models, having K-factors $>$3.9 (see Figure~\ref{fig_hist_ages_bar}). Although the two major peaks might be associated with the same cluster formation event, it is possible that the bar delayed its cluster formation with respect to the rest of the galaxy. Furthermore, the skewness of the outskirts distribution suggests a sudden termination of the cluster formation, contrary to the more continuous formation in the bar.

\begin{figure*}
\begin{center}
\includegraphics[scale=0.4]{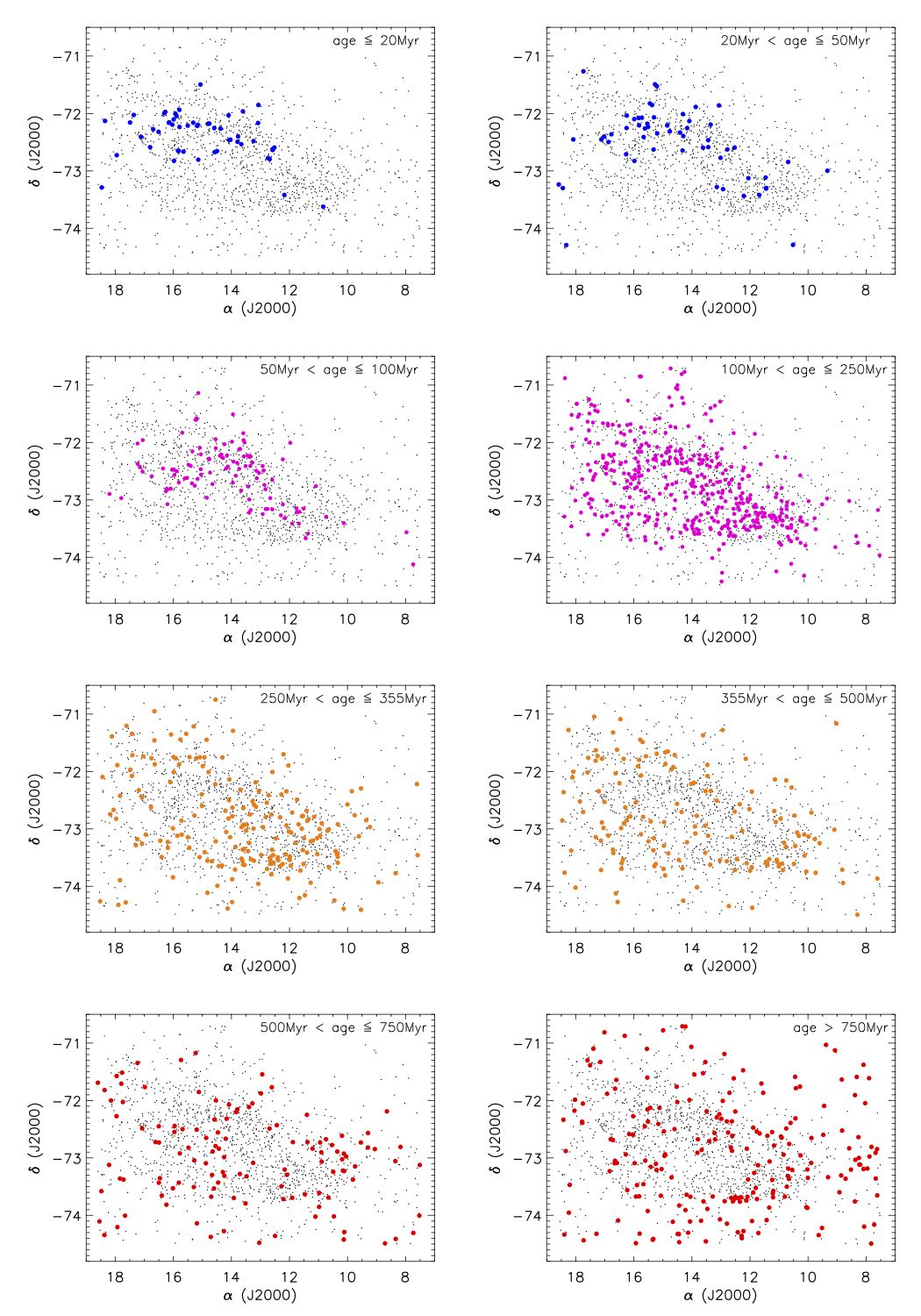}
\caption{Spatial age distribution for all the star clusters in our sample (black dots). The coordinates in both axes are in degrees (J2000). From top-left to bottom-right, we present the positions of star clusters with:  {\rm Age}$\le$20 Myr, 20$<${\rm Age}$\le$50 Myr, 50$<${\rm Age}$\le$100 Myr, 100$<${\rm Age}$\le$250 Myr, 250$<${\rm Age}$\le$355 Myr, 355$<${\rm Age}$\le$500 Myr, 500$<${\rm Age}$\le$750 Myr, and {\rm Age}$>$750 Myr. }
\label{fig_sfh}
\end{center}
\end{figure*}

The above results can be also confirmed from Figure~\ref{fig_sfh}, where we present the spatial distribution of clusters of different ages in our sample (the age ranges are as in \citealt{Bitsakis17}). Clusters younger than 100 Myr are solely located in the bar region, while clusters older than 355 Myr are mostly populating the outskirts. The bar is also associated with two prominent H{\rm I} supershells \citep{Stanimirovic99}, confirming the recent burst of star formation in that region. What is remarkable is the fact that, starting from the center of the SMC-bar, clusters of larger ages are gradually located outwards, with only very few old clusters ($>$750 Myr) found in the central region of the galaxy. This result suggests that an outside-in  quenching of cluster formation occurred over the past Gyr in the SMC.


\section{Discussion: Comparison between the LMC-SMC cluster ages and implications}
As presented above, our method is able to create complete, uniform samples of star clusters which allow comparisons between different galaxies. In particular, the use of an identical set-up and data as in \citet{Bitsakis17} secures the robustness of the comparisons between the star cluster properties of the two Magellanic Clouds, namely the SMC and LMC. 
 
\begin{figure*}
\begin{center}
\includegraphics[scale=0.32]{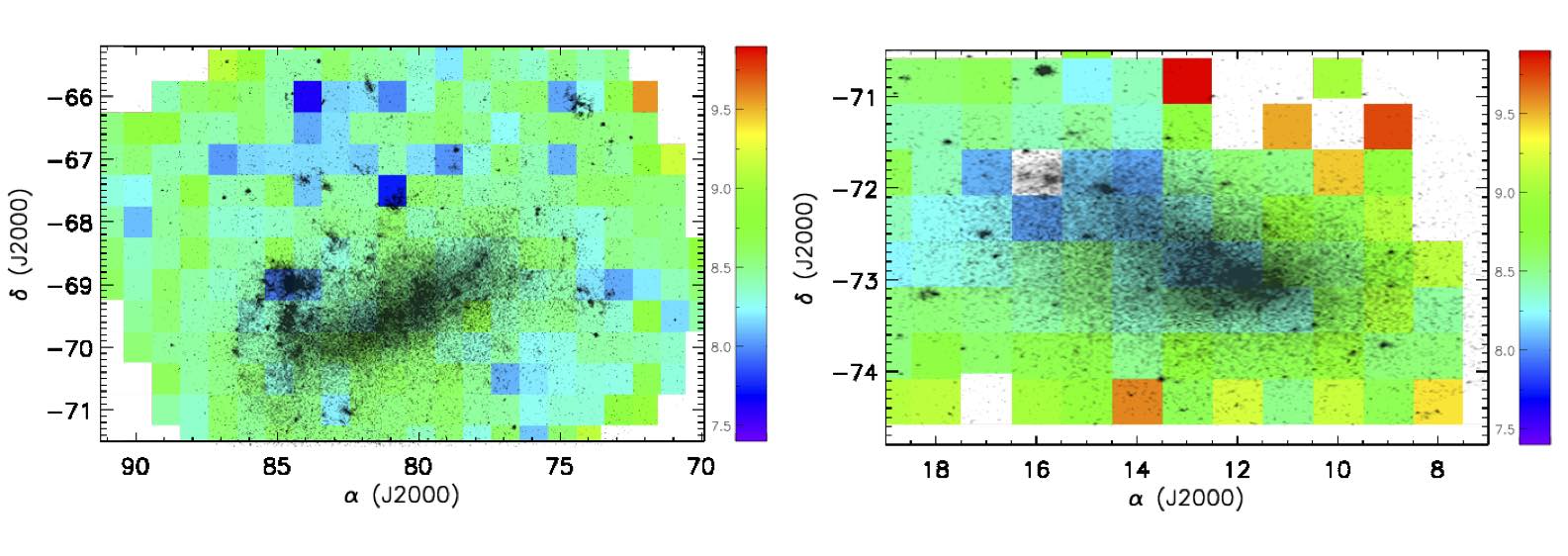}
\caption{The spatially binned median age distribution of the LMC (left) and the SMC (right), respectively, overlaid on their Spitzer/IRAC 3.6$\micron$ mosaics. The color-scale covers the range 7.4$\leq$log(age/yr)$<$9.9. The coordinates in both axis are in degrees (J2000). The bin sizes are 1 deg in RA and 0.5 deg in Dec.}
\label{fig_age_dens}
\end{center}
\end{figure*}

We compare the cluster age distributions of the two galaxies, presented in Figure~\ref{fig_hist_ages} of the current work for the SMC and in Figure~8 of \citet{Bitsakis17} for the LMC, and we discuss the implications. The comparison shows that both Clouds display enhanced cluster formation activity in the last 200-to-300 Myr. This is also consistent with the peaks of cluster formation in the bars of both galaxies; this age coincides with the epoch at which \citet{Besla12} estimated that a direct collision occurred between the two Clouds. Yet, owing to large differences in their sizes and masses, the effects of such a collision in the cluster formation history of the two galaxies should have been very different. This is evident in Figure~\ref{fig_age_dens}, where we present the median age distribution in bins $\sim$0.5 deg$^{2}$ for the LMC (left) and the SMC (right), respectively. It is shown that the star clusters in the SMC bar are younger than those in the LMC bar, where the most recent cluster formation occurred $>$50 Myr ago. In contrast, the SMC bar is experiencing an on-going cluster formation activity, with 8\% of its clusters (14\% of those located in the bar) having ages $<$50 Myr. This also agrees with the findings of \citet{Chiosi06} and \citet{Glatt10}, of very recent ($<$20 Myr) cluster formation activity in the SMC. This suggests the presence of cold molecular gas in the central region of that galaxy, as confirmed by \citet{Bolatto11}.

The age distributions of the outskirts of both galaxies also show great differences. The SMC contains on average clusters older than 300 Myr ($\sim$15\% of them are older than a Gyr), while the LMC contains mostly clusters 150-to-500 Myr old (only 7\% have ages $>$1 Gyr). Despite those differences, both distributions seem to have peaked $\sim$300 Myr ago, suggesting that the aforementioned collision between the two Clouds not only affected their bars, but rather triggered cluster formation on a global scale in those galaxies. The secondary SMC peak at 680 Myr might be matched with the smaller $\sim$500 Myr peak of the LMC. These results would then be in agreement with the 0.6 Gyr star formation enhancement observed by \citet{Harris04,Harris09}, who studied the star formation histories of the two galaxies and, based on orbital simulations available at the time, associated such events with perigalactic passages of the Magellanic Clouds about the Galaxy.

This difference in the old vs. young cluster spatial distributions suggests that the SMC may have ceased its star cluster formation in an outside-in fashion. This result is consistent with the findings of \citet{Cignoni13}, who studied the spatially resolved star formation history of six SMC regions and suggested the existence of an age gradient with all the star formation activity over the past 0.5 Gyr being concentrated in the central region. Such an age gradient has not been reported, however, for clusters older than 1 Gyr \citep[see][]{Parisi14}. This implies that its interaction with the LMC (or the Galaxy) could have affected (by stripping, shocks, or inflows towards the center) its \emph{outer} gas reservoir, thus preventing it from forming younger star clusters in the outskirts. \citet{Zhang12} studied the multi-band surface brightness profiles of 34 nearby dwarf irregular galaxies, and found an outside-in shrinking of the star formation that they attributed to environmental effects (i.e., interactions between galaxies). Arguably the LMC, being 50\% more massive than the SMC, did not suffer similar gas loss by galaxy-galaxy interactions, and hence retained its global cluster formation throughout its lifetime.

The comparison of  the spatial distributions of young clusters ($<$50 Myr) in the Magellanic Clouds is also puzzling. As shown in Figure~\ref{fig_age_dens}, clusters with these ages in the SMC are mostly located at the bar, preferentially at the bar-``arm''\footnote{The ``arms" are here intended as those HI features of the Magellanic Clouds resembling classical spiral arms, although their actual nature is still under debate, as described in the text.} junction points, while in the LMC they lie mostly along its arms. In the case of the LMC, HI arms are found north-east and south-west of the bar \citep{Kim03}, whereas in the SMC they trace an elongated structure located south-east of the bar \citep{Stanimirovic99, Dickey00}. Interestingly, \citet{Ochsendorf17} showed that the most active star forming regions at present in the LMC, namely the 30Dor and N79, are located where the LMC bar joins the HI arms. Such locations are very likely to enhance star formation due to the high concentrations of gas and to shocks induced by the internal dynamics, and very young stars/clusters have been observed there in various other galaxies \citep[e.g.,][]{Beuther17}.  The absence of young clusters in the outskirts of the SMC is likely due to the overall scarcity of gas in the last few Myr. The hypothesis of outside-in stripping of the gas in the SMC is also consistent with the cold molecular gas distribution \citep[see][]{Bolatto11}. In the SMC, molecular gas is mostly confined to the bar, and indeed its youngest clusters overlap the densest molecular gas in the north-eastern portion of the bar at its intersection with the aforementioned HI feature. Using the Spitzer/MIPS 24$\micron$ images we confirm that the locations of the young clusters coincide with those of the warm dust clouds too. It is plausible that many of those clusters are still embedded in the progenitor clouds, thus explaining their very young ages. 

Regarding the LMC, since many of the clusters younger than 50 Myr seem to trace both HI arms \citep{Bitsakis17}, we have considered the possibility that star formation there is related to a long-lived spiral density wave not connected to an interaction with the SMC. This hypothesis, however, is disproven by the absence of a corresponding density enhancement in the old stellar disk as traced by the near-IR, as reported by \citet{Marel01} and confirmed by our own multiwavelength analysis.  On the other hand, if the LMC bar was excited or enhanced by an interaction with the SMC a few Myr ago, the present-day star formation in the LMC should still be traced back to that interaction, if indeed star formation is triggered by shocks in the bar-arm interface, especially when the pattern speeds of bar and arms are different \citep{Beuther17,MartinezGarcia11}. 

Our results suggest that, in spite of the asymmetries in the cluster formation histories of the two galaxies, their overall evolution is a combination of both internal and environmental mechanisms. \citet{Harris04, Harris09} suggested that the star formation histories of the Clouds are dominated by correlated -- thus environmental -- mechanisms. Our findings agree with their conclusions that the interactions between the Magellanic Clouds and the Galaxy were predominant in shaping uniquely the star cluster formation history in the Clouds.

\section{Conclusions}
We applied our new method to detect and estimate the ages of star clusters in nearby galaxies \citep[originally presented in][]{Bitsakis17} on the multi-band, high resolution data of the SMC. We apply the same set-up and procedure to analogous data of the two galaxies, and compare the results. Our conclusions are summarized below.

\begin{enumerate}

\item[(a)] We detect 1319 star clusters in the central 18 deg$^{2}$ of the SMC we surveyed. 1108 of these clusters have never been reported before.

\item[(b)] The distribution of cluster ages suggests major star cluster formation $\sim$240 Myr ago. Studying the corresponding distributions of the SMC bar and outskirts, we find that they have significant differences, with the cluster formation peaking at the bar $\sim$200 Myr ago, while for the rest of the galaxy the average age is $\sim$270 Myr ago. Moreover, the skewness of the age distribution in the galaxy outskirts suggests a termination of the cluster formation over the past few Myr.

\item[(c)] The spatially resolved age distribution of the star clusters in the SMC suggests that the inner part of the galaxy was formed more recently, and that an outside-in quenching of cluster formation occurred over the past Gyr. 

\item[(d)] A comparison between the above results and those derived previously for the LMC shows that both galaxies have experienced an intense star cluster formation event at $\sim$300 Myr ago, consistent with a direct collision scenario proposed by model simulations.   

\item[(e)] Most of the youngest clusters in both Magellanic Clouds are found where their bars meet the HI arms (or similar elongated features), suggesting that cluster formation there is triggered by internal dynamical processes. 

\item[(f)] Our results suggest that the interactions between Magellanic Clouds are the major driver of their large-scale star cluster formation and overall evolution.

\end{enumerate}

\acknowledgments

The authors wish to thank the anonymous referee for her/his thorough review and valuable comments that helped improve significantly this article. TB would like to acknowledge support from the CONACyT Research Fellowships program. We gratefully acknowledge support from the program for basic research of CONACyT through grant number 252364. GM acknowledges support from CONICYT, Programa de Astronom\'ia/PCI, FONDO ALMA 2014, Proyecto No 31140024. GB acknowledges support for this work from UNAM through grant PAPIIT IG100115. This research made use of TOPCAT, an interactive graphical viewer and editor for tabular data. IRAF is distributed by the National Optical Astronomy Observatory, which is operated by the Association of Universities for Research in Astronomy (AURA) under cooperative agreement with the National Science Foundation.

\bibliography{LMCbib.bib} 

\begin{thebibliography}{}
\expandafter\ifx\csname natexlab\endcsname\relax\def\natexlab#1{#1}\fi
\providecommand{\url}[1]{\href{#1}{#1}}

\bibitem[{{Baumgardt} {et~al.}(2013){Baumgardt}, {Parmentier}, {Anders}, \&
  {Grebel}}]{Baumgardt13}
{Baumgardt}, H., {Parmentier}, G., {Anders}, P., \& {Grebel}, E.~K. 2013,
  \mnras, 430, 676

\bibitem[{{Besla} {et~al.}(2007){Besla}, {Kallivayalil}, {Hernquist},
  {Robertson}, {Cox}, {van der Marel}, \& {Alcock}}]{Besla07}
{Besla}, G., {Kallivayalil}, N., {Hernquist}, L., {et~al.} 2007, \apj, 668, 949

\bibitem[{{Besla} {et~al.}(2012){Besla}, {Kallivayalil}, {Hernquist}, {van der
  Marel}, {Cox}, \& {Kere{\v s}}}]{Besla12}
---. 2012, \mnras, 421, 2109

\bibitem[{{Beuther} {et~al.}(2017){Beuther}, {Meidt}, {Schinnerer}, {Paladino},
  \& {Leroy}}]{Beuther17}
{Beuther}, H., {Meidt}, S., {Schinnerer}, E., {Paladino}, R., \& {Leroy}, A.
  2017, \aap, 597, A85

\bibitem[{{Bica} {et~al.}(2008){Bica}, {Bonatto}, {Dutra}, \&
  {Santos}}]{Bica08}
{Bica}, E., {Bonatto}, C., {Dutra}, C.~M., \& {Santos}, J.~F.~C. 2008, \mnras,
  389, 678

\bibitem[{{Bitsakis} {et~al.}(2017){Bitsakis}, {Bonfini},
  {Gonz{\'a}lez-L{\'o}pezlira}, {Ram{\'{\i}}rez-Siordia}, {Bruzual}, {Charlot},
  {Maravelias}, \& {Zaritsky}}]{Bitsakis17}
{Bitsakis}, T., {Bonfini}, P., {Gonz{\'a}lez-L{\'o}pezlira}, R.~A., {et~al.}
  2017, \apj, 845, 56

\bibitem[{{Bolatto} {et~al.}(2011){Bolatto}, {Leroy}, {Jameson}, {Ostriker},
  {Gordon}, {Lawton}, {Stanimirovi{\'c}}, {Israel}, {Madden}, {Hony},
  {Sandstrom}, {Bot}, {Rubio}, {Winkler}, {Roman-Duval}, {van Loon},
  {Oliveira}, \& {Indebetouw}}]{Bolatto11}
{Bolatto}, A.~D., {Leroy}, A.~K., {Jameson}, K., {et~al.} 2011, \apj, 741, 12

\bibitem[{{Boutloukos} \& {Lamers}(2003)}]{Boutloukos03}
{Boutloukos}, S.~G., \& {Lamers}, H.~J.~G.~L.~M. 2003, \mnras, 338, 717

\bibitem[{{Chen} {et~al.}(2015){Chen}, {Bressan}, {Girardi}, {Marigo}, {Kong},
  \& {Lanza}}]{Chen15}
{Chen}, Y., {Bressan}, A., {Girardi}, L., {et~al.} 2015, \mnras, 452, 1068

\bibitem[{{Chiosi} {et~al.}(2006){Chiosi}, {Vallenari}, {Held}, {Rizzi}, \&
  {Moretti}}]{Chiosi06}
{Chiosi}, E., {Vallenari}, A., {Held}, E.~V., {Rizzi}, L., \& {Moretti}, A.
  2006, \aap, 452, 179

\bibitem[{{Cignoni} {et~al.}(2013){Cignoni}, {Cole}, {Tosi}, {Gallagher},
  {Sabbi}, {Anderson}, {Grebel}, \& {Nota}}]{Cignoni13}
{Cignoni}, M., {Cole}, A.~A., {Tosi}, M., {et~al.} 2013, \apj, 775, 83

\bibitem[{{de Grijs} \& {Goodwin}(2009)}]{deGrijs09}
{de Grijs}, R., \& {Goodwin}, S.~P. 2009, in IAU Symposium, Vol. 256, The
  Magellanic System: Stars, Gas, and Galaxies, ed. J.~T. {Van Loon} \& J.~M.
  {Oliveira}, 311--316

\bibitem[{{Dickey} {et~al.}(2000){Dickey}, {Mebold}, {Stanimirovic}, \&
  {Staveley-Smith}}]{Dickey00}
{Dickey}, J.~M., {Mebold}, U., {Stanimirovic}, S., \& {Staveley-Smith}, L.
  2000, \apj, 536, 756

\bibitem[{{Fazio} {et~al.}(2004){Fazio}, {Hora}, {Allen}, {Ashby}, {Barmby},
  {Deutsch}, {Huang}, {Kleiner}, {Marengo}, {Megeath}, {Melnick}, {Pahre},
  {Patten}, {Polizotti}, {Smith}, {Taylor}, {Wang}, {Willner}, {Hoffmann},
  {Pipher}, {Forrest}, {McMurty}, {McCreight}, {McKelvey}, {McMurray}, {Koch},
  {Moseley}, {Arendt}, {Mentzell}, {Marx}, {Losch}, {Mayman}, {Eichhorn},
  {Krebs}, {Jhabvala}, {Gezari}, {Fixsen}, {Flores}, {Shakoorzadeh}, {Jungo},
  {Hakun}, {Workman}, {Karpati}, {Kichak}, {Whitley}, {Mann}, {Tollestrup},
  {Eisenhardt}, {Stern}, {Gorjian}, {Bhattacharya}, {Carey}, {Nelson},
  {Glaccum}, {Lacy}, {Lowrance}, {Laine}, {Reach}, {Stauffer}, {Surace},
  {Wilson}, {Wright}, {Hoffman}, {Domingo}, \& {Cohen}}]{Fazio04}
{Fazio}, G.~G., {Hora}, J.~L., {Allen}, L.~E., {et~al.} 2004, \apjs, 154, 10

\bibitem[{{Glatt} {et~al.}(2010){Glatt}, {Grebel}, \& {Koch}}]{Glatt10}
{Glatt}, K., {Grebel}, E.~K., \& {Koch}, A. 2010, \aap, 517, A50

\bibitem[{{Gordon} {et~al.}(2011){Gordon}, {Meixner}, {Meade}, {Whitney},
  {Engelbracht}, {Bot}, {Boyer}, {Lawton}, {Sewi{\l}o}, {Babler}, {Bernard},
  {Bracker}, {Block}, {Blum}, {Bolatto}, {Bonanos}, {Harris}, {Hora},
  {Indebetouw}, {Misselt}, {Reach}, {Shiao}, {Tielens}, {Carlson},
  {Churchwell}, {Clayton}, {Chen}, {Cohen}, {Fukui}, {Gorjian}, {Hony},
  {Israel}, {Kawamura}, {Kemper}, {Leroy}, {Li}, {Madden}, {Marble},
  {McDonald}, {Mizuno}, {Mizuno}, {Muller}, {Oliveira}, {Olsen}, {Onishi},
  {Paladini}, {Paradis}, {Points}, {Robitaille}, {Rubin}, {Sandstrom}, {Sato},
  {Shibai}, {Simon}, {Smith}, {Srinivasan}, {Vijh}, {Van Dyk}, {van Loon}, \&
  {Zaritsky}}]{Gordon11}
{Gordon}, K.~D., {Meixner}, M., {Meade}, M.~R., {et~al.} 2011, \aj, 142, 102

\bibitem[{{Harris}(2007)}]{Harris07}
{Harris}, J. 2007, \apj, 658, 345

\bibitem[{{Harris} \& {Zaritsky}(2004)}]{Harris04}
{Harris}, J., \& {Zaritsky}, D. 2004, \aj, 127, 1531

\bibitem[{{Harris} \& {Zaritsky}(2009)}]{Harris09}
---. 2009, \aj, 138, 1243

\bibitem[{{Hilditch} {et~al.}(2005){Hilditch}, {Howarth}, \&
  {Harries}}]{Hilditch05}
{Hilditch}, R.~W., {Howarth}, I.~D., \& {Harries}, T.~J. 2005, \mnras, 357, 304

\bibitem[{{Kallivayalil} {et~al.}(2013){Kallivayalil}, {van der Marel},
  {Besla}, {Anderson}, \& {Alcock}}]{Kallivayalil13}
{Kallivayalil}, N., {van der Marel}, R.~P., {Besla}, G., {Anderson}, J., \&
  {Alcock}, C. 2013, \apj, 764, 161

\bibitem[{{Kim} {et~al.}(2003){Kim}, {Staveley-Smith}, {Dopita}, {Sault},
  {Freeman}, {Lee}, \& {Chu}}]{Kim03}
{Kim}, S., {Staveley-Smith}, L., {Dopita}, M.~A., {et~al.} 2003, \apjs, 148,
  473

\bibitem[{{Lejeune} {et~al.}(1997){Lejeune}, {Cuisinier}, \&
  {Buser}}]{Lejeune97}
{Lejeune}, T., {Cuisinier}, F., \& {Buser}, R. 1997, \aaps, 125, 229

\bibitem[{{Marigo} {et~al.}(2013){Marigo}, {Bressan}, {Nanni}, {Girardi}, \&
  {Pumo}}]{Marigo13}
{Marigo}, P., {Bressan}, A., {Nanni}, A., {Girardi}, L., \& {Pumo}, M.~L. 2013,
  \mnras, 434, 488

\bibitem[{{Martin} {et~al.}(2005){Martin}, {Fanson}, {Schiminovich},
  {Morrissey}, {Friedman}, {Barlow}, {Conrow}, {Grange}, {Jelinsky},
  {Milliard}, {Siegmund}, {Bianchi}, {Byun}, {Donas}, {Forster}, {Heckman},
  {Lee}, {Madore}, {Malina}, {Neff}, {Rich}, {Small}, {Surber}, {Szalay},
  {Welsh}, \& {Wyder}}]{Martin05}
{Martin}, D.~C., {Fanson}, J., {Schiminovich}, D., {et~al.} 2005, \apjl, 619,
  L1

\bibitem[{{Mart{\'{\i}}nez-Garc{\'{\i}}a} \&
  {Gonz{\'a}lez-L{\'o}pezlira}(2011)}]{MartinezGarcia11}
{Mart{\'{\i}}nez-Garc{\'{\i}}a}, E.~E., \& {Gonz{\'a}lez-L{\'o}pezlira}, R.~A.
  2011, \apj, 734, 122

\bibitem[{{Mighell} {et~al.}(1996){Mighell}, {Rich}, {Shara}, \&
  {Fall}}]{Mighell96}
{Mighell}, K.~J., {Rich}, R.~M., {Shara}, M., \& {Fall}, S.~M. 1996, \aj, 111,
  2314

\bibitem[{{Ochsendorf} {et~al.}(2017){Ochsendorf}, {Zinnecker}, {Nayak},
  {Bally}, {Meixner}, {Jones}, {Indebetouw}, \& {Rahman}}]{Ochsendorf17}
{Ochsendorf}, B.~B., {Zinnecker}, H., {Nayak}, O., {et~al.} 2017, Nature
  Astronomy, 1, 268

\bibitem[{{Olsen} {et~al.}(2011){Olsen}, {Zaritsky}, {Blum}, {Boyer}, \&
  {Gordon}}]{Olsen11}
{Olsen}, K.~A.~G., {Zaritsky}, D., {Blum}, R.~D., {Boyer}, M.~L., \& {Gordon},
  K.~D. 2011, \apj, 737, 29

\bibitem[{{Parisi} {et~al.}(2014){Parisi}, {Geisler}, {Carraro}, {Clari{\'a}},
  {Costa}, {Grocholski}, {Sarajedini}, {Leiton}, \& {Piatti}}]{Parisi14}
{Parisi}, M.~C., {Geisler}, D., {Carraro}, G., {et~al.} 2014, \aj, 147, 71

\bibitem[{{Rafelski} \& {Zaritsky}(2005)}]{Rafelski05}
{Rafelski}, M., \& {Zaritsky}, D. 2005, \aj, 129, 2701

\bibitem[{Richardson \& Green(1997)}]{Richardson97}
Richardson, S., \& Green, P.~J. 1997, Journal of the Royal Statistical Society,
  59, 731.
\newblock \url{http://dx.doi.org/10.1111/1467-9868.00095}

\bibitem[{{Rieke} {et~al.}(2004){Rieke}, {Young}, {Engelbracht}, {Kelly},
  {Low}, {Haller}, {Beeman}, {Gordon}, {Stansberry}, {Misselt}, {Cadien},
  {Morrison}, {Rivlis}, {Latter}, {Noriega-Crespo}, {Padgett}, {Stapelfeldt},
  {Hines}, {Egami}, {Muzerolle}, {Alonso-Herrero}, {Blaylock}, {Dole}, {Hinz},
  {Le Floc'h}, {Papovich}, {P{\'e}rez-Gonz{\'a}lez}, {Smith}, {Su}, {Bennett},
  {Frayer}, {Henderson}, {Lu}, {Masci}, {Pesenson}, {Rebull}, {Rho}, {Keene},
  {Stolovy}, {Wachter}, {Wheaton}, {Werner}, \& {Richards}}]{Rieke04}
{Rieke}, G.~H., {Young}, E.~T., {Engelbracht}, C.~W., {et~al.} 2004, \apjs,
  154, 25

\bibitem[{{Schmeja}(2011)}]{Schemja10}
{Schmeja}, S. 2011, Astronomische Nachrichten, 332, 172

\bibitem[{{Siegel} {et~al.}(2014){Siegel}, {Porterfield}, {Linevsky}, {Bond},
  {Holland}, {Hoversten}, {Berrier}, {Breeveld}, {Brown}, \&
  {Gronwall}}]{Siegel14}
{Siegel}, M.~H., {Porterfield}, B.~L., {Linevsky}, J.~S., {et~al.} 2014, \aj,
  148, 131

\bibitem[{{Simons} {et~al.}(2014){Simons}, {Thilker}, {Bianchi}, \&
  {Wyder}}]{Simons14}
{Simons}, R., {Thilker}, D., {Bianchi}, L., \& {Wyder}, T. 2014, Advances in
  Space Research, 53, 939

\bibitem[{{Stanimirovic} {et~al.}(1999){Stanimirovic}, {Staveley-Smith},
  {Dickey}, {Sault}, \& {Snowden}}]{Stanimirovic99}
{Stanimirovic}, S., {Staveley-Smith}, L., {Dickey}, J.~M., {Sault}, R.~J., \&
  {Snowden}, S.~L. 1999, \mnras, 302, 417

\bibitem[{{Stetson}(1987)}]{Stetson87}
{Stetson}, P.~B. 1987, \pasp, 99, 191

\bibitem[{{van der Marel}(2001)}]{Marel01}
{van der Marel}, R.~P. 2001, \aj, 122, 1827

\bibitem[{{Venn}(1999)}]{Venn99}
{Venn}, K.~A. 1999, \apj, 518, 405

\bibitem[{{Yoshizawa} \& {Noguchi}(2003)}]{Yoshizawa03}
{Yoshizawa}, A.~M., \& {Noguchi}, M. 2003, \mnras, 339, 1135

\bibitem[{{Zaritsky} {et~al.}(2002){Zaritsky}, {Harris}, {Thompson}, {Grebel},
  \& {Massey}}]{Zaritsky02}
{Zaritsky}, D., {Harris}, J., {Thompson}, I.~B., {Grebel}, E.~K., \& {Massey},
  P. 2002, \aj, 123, 855

\bibitem[{{Zhang} {et~al.}(2012){Zhang}, {Hunter}, {Elmegreen}, {Gao}, \&
  {Schruba}}]{Zhang12}
{Zhang}, H.-X., {Hunter}, D.~A., {Elmegreen}, B.~G., {Gao}, Y., \& {Schruba},
  A. 2012, \aj, 143, 47

\end{thebibliography}




\end{document}